\documentclass[conference]{IEEEtran}
\IEEEoverridecommandlockouts
\usepackage{cite}
\usepackage{graphicx}
\usepackage{listings}
\def\BibTeX{{\rm B\kern-.05em{\sc i\kern-.025em b}\kern-.08em
    T\kern-.1667em\lower.7ex\hbox{E}\kern-.125emX}}
\begin{document}

\title{OAuth 2.0 authorization using blockchain-based tokens}
\author{\IEEEauthorblockN{Nikos Fotiou, Iakovos Pittaras, Vasilios A. Siris, Spyros Voulgaris, George C. Polyzos}
\IEEEauthorblockA{Mobile Multimedia Laboratory,\\
Department of Informatics School of Information Sciences and Technology\\
Athens University of Economics and Business, Greece\\
\{fotiou,pittaras,vsiris,voulgaris,polyzos\}@aueb.gr}
}

\maketitle

\begin{abstract}
    OAuth~2.0 is the industry-standard protocol for authorization. It facilitates secure service provisioning, as well as secure interoperability among diverse stakeholders. All OAuth~2.0 protocol flows result in the creation of an access token, which is then used by a user to request access to a protected resource. Nevertheless, the definition of access tokens is transparent to the OAuth~2.0 protocol, which does not specify any particular token format, how tokens are generated, or how they are used. Instead, the OAuth 2.0 specification leaves all these as design choices for integrators. In this paper, we propose a new type of OAuth~2.0 token backed by a distributed ledger. Our construction is secure, and it supports proof-of-possession, auditing, and accountability. Furthermore, we provide added-value token management services, including revocation, delegation, and fair exchange by leveraging smart contracts. We realized a proof-of-concept implementation of our solution using Ethereum smart contracts and the ERC-721 token specification. 
\end{abstract}

\section{Introduction}
OAuth 2.0~\cite{oauth} has received widespread adoption and it is considered the industry-standard protocol for authorization. OAuth~2.0 enables delegation and interoperability, it enhances end-user security, and it facilitates access control management. Due to its intriguing properties, it is being used in environments with higher security requirements than initially considered, such as IoT systems, Open Banking, eHealth, eGovernment, and Electronic Signatures~\cite{Lod2019}.

In a nutshell, OAuth 2.0 constitutes the protocol through which an 
\emph{client} obtains an \emph{access token} from an \emph{authorization server}, to access a protected resource, stored in a \emph{resource server}. However, OAuth~2.0  specification does not define how an access token is generated, validated, and destroyed; instead it leaves the management of OAuth~2.0 tokens' lifecycle as an open design choice.       

In this paper, we propose a new type of token, which is protected using proof-of-possession keys and at the same time it supports auditing and accountability, fast revocation, and added-value services. In order to achieve our goal, we build on the emerging distributed ledger technology (DLT). Our solution considers an append-only distributed ledger (public or private), where users, identified by a public key, can transact \emph{uniquely identified tokens}. Our implementation is based on the Ethereum blockchain and the ERC-721 token specification (but other similar technologies can be considered). Our system leverage Ethereum's support for distributed apps (referred to as \emph{smart contracts}) to build blockchain-based token management services. The proposed solution has the following advantages:
 \begin{itemize}
    \item The entity that generates access tokens (i.e., the authorization server) can easily revoke them before they expire. Furthermore, token revocation does not involve any interaction with the client or the resource server, hence it can be implemented even if these entities are offline/unreachable. Similarly, a resource server can verify the validity of a token (i.e., it has not been revoked) even if the authorization server is offline, since token revocation is recorded on the DLT through a transaction that can occur at any time prior to the token's usage. 
    \item Clients do not have to store their tokens locally, neither do they have to store any secret associated with their access tokens. Instead, all tokens can be retrieved form the ledger. Hence, tokens are portable and can be easily used by multiple client devices. Furthermore, since we are using a popular token specification, a wide range of ``wallets'' and libraries can be supported by our system.
    \item Token integrity and authenticity can be verified simply by performing a lookup in the ledger, and it does not involve any signature verification or any other cryptographic operation. Therefore, our solution is less prone to implementation errors, and tokens are simpler since the do not carry any cryptographic proof. Furthermore, token ownership can be securely modified without any interaction with the authorization server.
    \item All tokens, including the revoked ones, are immutably stored in the ledger, hence auditing and accountability mechanisms are facilitated. 
 \end{itemize}
 
The feasibility of our solution is verified through a proof-of-concept implementation, inspired by an IoT gateway access use case. The remainder of this paper is organized as follows. In Section~\ref{sec:back}, we introduce OAuth~2.0 and ERC-721 tokens. In Section~\ref{sec:sys} we present the design of our system, and in Section~\ref{sec:tok} we detail the provided token management services. Finally, we present our implementation and its evaluation in Section~\ref{sec:eval}, and we conclude in Section~\ref{sec:conc}. 
 
 \section{Background and related work}
 \label{sec:back}
 \subsection{OAuth 2.0}
 \label{sec:oauth}
\begin{figure}
    \centering
    \includegraphics[width=0.8\linewidth]{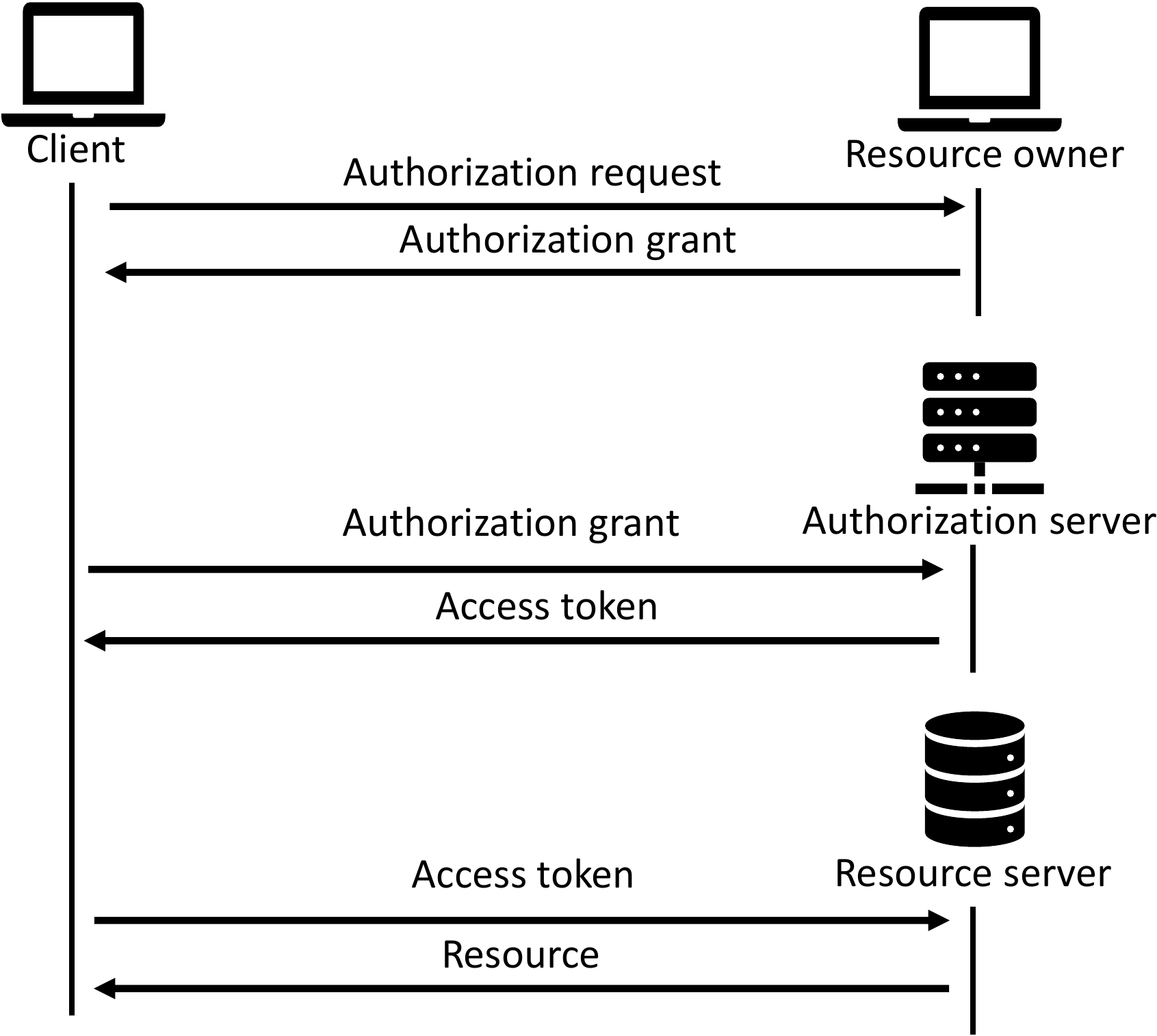}
    \caption {OAuth~2.0 interactions.}
    \label{fig:oauth}
\end{figure}

 OAuth 2.0 systems are composed of the following entities. A \emph{resource server} that hosts a protected resource owned by a \emph{resource owner}, a \emph{client} wishing to access that resource, and an \emph{authorization server} responsible for generating \emph{access tokens}. Access tokens are granted to clients authorized by the resource owner: client authorization is proven using an \emph{authorization grant}. These interactions are illustrated in Fig.~\ref{fig:oauth}. As it can be seen in this figure, a client first requests an authorization grant from the resource owner, then it uses this grant to obtain an access token from the authorization server, and finally, it uses the access token to access the protected resource stored in the resource server. The semantics, as well as the mechanisms for generating and validating access tokens and grants are transparent to the OAuth~2.0 protocol.  

 Each OAuth~2.0 deployment may choose the type of token it will use. The most commonly used type of token is the \emph{bearer} token~\cite{Jon2012}, which can be used by any user who possesses it (i.e., the ``bearer'').  For additional security, a token can be associated with a secret key, so that only users who can prove that they possess this key can use the token. Since the latter type of tokens provides more security (at the cost of the communication overhead required to verify ownership) it is considered by our solution. In particular, our constructions are based on JSON Web Tokens (JWTs)~\cite{Jon2015} enhanced with blockchain-based proof-of-possession mechanisms. 

 A JWT is a compact, URL-safe means of representing ``claims.'' It consists of  zero or more name/value pairs, and it is transmitted encoded in base64url~\cite{rfc4648}. RFC 7519 defines some ``standard'' claim names and their semantics. Table~\ref{table:jwt} contains the names, and the corresponding semantics, of the JWT claims used by our solution. 

 \begin{table}[h]
    \centering
    \caption{JWT claims used in our system}
    \begin{tabular}{ | l | p{3cm} |}
    \hline
    \textbf{Name} & \textbf{Semantics} \\ \hline
       \texttt{iss} &  The issuer of the token \\ \hline
       \texttt{sub} &  The subject of the token, i.e., the entity that will use the token to gain access to a resource \\ \hline
       \texttt{aud} &  The audience of the token, i.e., the the recipients that the JWT is intended for  \\ \hline
       \texttt{exp} &  The expiration time on or after which the JWT must not be accepted for processing \\ \hline
       \texttt{jti} &  A unique token identifier \\ \hline
    \end{tabular}
 
    \label{table:jwt}
\end{table}


\subsection{Ethereum and ERC-721}
\label{sec:eth}
Ethereum~\cite{Wood2014} is a popular blockchain system that supports distributed applications, known as ``smart contracts''. A smart contract is executed by all ``peers'' in the Ethereum network and its outcome is agreed upon consensus. Users interact with a smart contract by issuing ``transactions,'' which are signed by a user-specific key. A hash of this key is used as the user ``address'' and it is used for associating users with information stored in the blockchain. All transactions are immutably recorded in the blockchain, for this reason blockchain-based solutions are ideal for implementing auditing and accountability mechanisms. Furthermore, a smart contract can create an ``event''. Events are also recorded in the blockchain and end-user applications can be configured to ``listen'' for specific contract events. 

Ethereum community is developing ``Ethereum Request for Comments'' (ERC), which the equivalent of RFC but for smart contract.   
ERC-721\footnote{https://eips.ethereum.org/EIPS/eip-721}, is an open standard that describes how to build ``non-fungible or unique tokens on the Ethereum blockchain.'' This standard is very similar, in many ways, to ERC-20\footnote{https://eips.ethereum.org/EIPS/eip-20}, which probably the most popular Ethereum standard and it is used for creating custom Ethereum tokens. However, in contrast to ERC-20 tokens, ERC-721 tokens are ``unique'' and non-interchangeable with other tokens (non-fungibility). Many Ethereum wallets, such as Metamask\footnote{https://metamask.io/}, can handle these tokens. All ERC-721-based tokens are identified by a unique identifier (we will refer to this identifier as $token_{id}$), and they can be owned only by a single user.  

This standard, like every other token standard in Ethereum, defines some functions that a smart contract should implement in order to be able to create and handle ERC-721 tokens. 
Furthermore, the ERC-721 metadata extension, defines some additional functions that can be used for associating an ERC-721 token with metadata. 
Table~\ref{table:erc} describes the functions defined in ERC-721 and in ERC-721 metadata extension, used by our system.

\begin{table}[h]
    \centering
    \caption{ERC-721 and ERC-721 metadata extension functions used in our system}
    \begin{tabular}{ | l | p{3cm} |}
    \hline
    \textbf{Name} & \textbf{Purpose} \\ \hline
      \texttt{ownerOf($token_{id}$)} &  It accepts as input a $token_{id}$ and returns the address of the token owner \\ \hline
      \texttt{transferFrom(from, to, $token_{id}$)} &  It transfers a  $token_{id}$ from one Ethereum address to another\\ \hline
      \texttt{approve(address,$token_{id}$)} &  It approves an Ethereum address to manage a $token_{id}$ on owner's behalf \\ \hline
      \texttt{getApproved($token_{id}$)} &  It retrieves the Ethereum address that is allowed to manage $token_{id}$\\ \hline
      \texttt{tokenURI($token_{id}$)} &  It accepts as input a $token_{id}$ and returns a URI that point to token's metadata \\ \hline
     \end{tabular}
 
    \label{table:erc}
\end{table}

Additionally,  functions \texttt{transferFrom} and \texttt{approve}, when invoked each generates an event, named \texttt{Transfer} and \texttt{Approval} respectively. Both these events have three attributes; the attributes of the \texttt{Transfer} event are the $from$ address, the $to$ address, and the $token_{id}$, while the attributes of the \texttt{Approval} event are the owner address, the approved address, and the $token_{id}$.  

%
\subsection{Related work}
Many recent research efforts investigate Blockchain-based access control, either by defining custom blockchain systems (e.g., \cite{Dor2017}) or by using Ethereum smart contracts for recording policies and for implementing on-chain access control decisions (e.g., \cite{Ham2018, Nov2018, Zha2019,Ali2019, Mae2019}). Our solution does not rely on smart contracts for implementing access control decisions (which may cause privacy issues), instead it uses the ledger for storing auxiliary information used for validating OAuth~2.0 access tokens. 

The work in this paper is related to our previous work published in~\cite{Fot2018,Lag2019,Sir2019}. However, in these papers we considered constrained devices not capable of accessing the blockchain. In this paper we relax this assumption and we consider resource servers capable of reading values from the blockchain (resource servers are not required to ``write'' in the blockchain); even with read-only access we can implement constructions that were not possible with our previous work, including revocation, and delegation. 

\section{System design}
\label{sec:sys}
\begin{figure}
    \centering
    \includegraphics[width=0.8\linewidth]{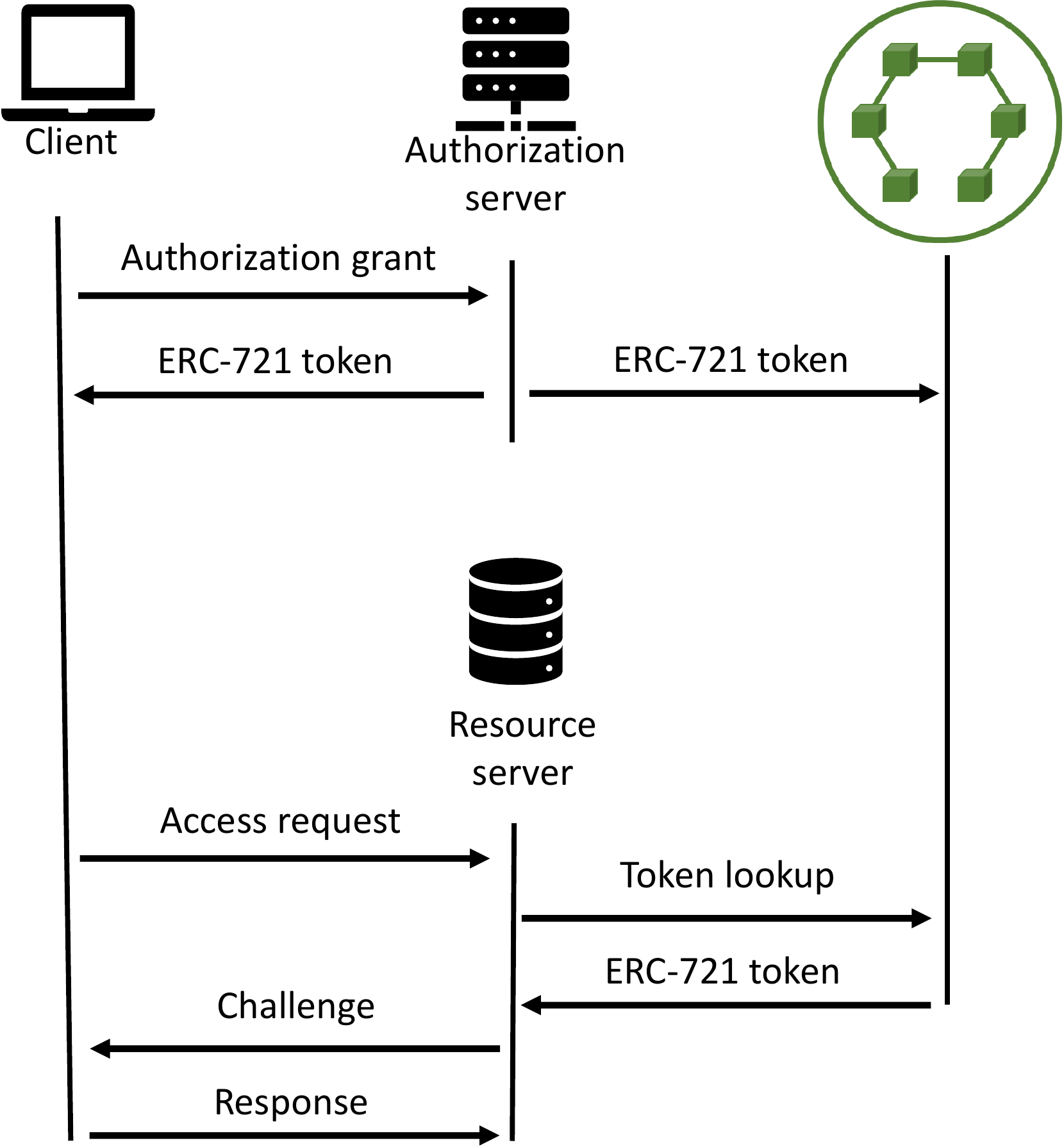}
    \caption {Overview of our system}
    \label{fig:overview}
\end{figure}

Our system considers a typical OAuth~2.0 architecture (such as the one described in section~\ref{sec:oauth}). Therefore, the main entities of our system are \emph{clients}, \emph{authorization servers}, \emph{resource servers}, and \emph{resource owners}.  From a high-level perspective, our system operates as follows. The client requests an access token from the authorization server, using an authorization grant received by the resource owner. The authorization server validates the authorization grant of the client, generates a JWT and an ERC-721 token, transfers the ERC-721 token to the Ethereum address of the client, and transmits the JWT to the client. Then, the client requests access from the resource server, providing the JWT. The resource server retrieves the corresponding ERC-721 token which is used for verifying the validity and the ownership of the JWT: if all verifications are successful the resource server allows the client request. This process is illustrated in Figure~\ref{fig:overview}.

\subsection{Notation and security assumptions}
In our system, resources are uniquely identified by a URI, referred to as the $URI_{resource}$. We assume a security mechanism with which a resource server can prove that it hosts a specific $URI_{resource}$ (e.g., using an X.509 certificate). Henceforth, when we say that ``a client requests $URI_{resource}$ from a resource server'' it is assumed that the client has already verified $URI_{resource}$ ownership. Moreover, the communication between a client and a resource server takes place over a secure communication channel. 

In our system each authorization server owns an ERC-721 contract, referred to as $Contract_{AS}$; we detail these smart contracts in the following section. Moreover, we assume a security mechanism
 with which clients and authorization servers can securely communicate; this mechanism provides confidentiality and integrity protection of the exchanged messages, as well as authorization server authentication. Furthermore, resource owners trust authorization servers to behave according to the specified procedures. 

Finally, each client owns a public key $P_{Client}$, and a corresponding private key, used for transacting with the Ethereum blockchain. Again, we consider a security mechanism with which $P_{Client}$ ownership can be verified. 

\subsection{The ERC-721 contract}
ERC-721 contracts in our system are implemented such that only the contract owner can create new tokens. When a token is created its $token_{id}$ and $metadata$ are specified: these two properties are read-only and they cannot be modified. Moreover, token owners in our system \emph{cannot transfer their tokens}; the only entity that can invoke the \texttt{transferFrom} method is the contract owner. Moreover, when invoking the \texttt{transferFrom} method, the contract owner is allowed to transfer any token, no matter who the token owner is. 

The ERC-recommended approach for associating a token with some metadata is through the metadata extension, which provides a method that maps a $token_{id}$ to a URI where metadata are stored (i.e., \texttt{tokenURI}). Nevertheless, we postulate that this approach violates both the decentralization and immutability principles of DLTs, since with this approach the metadata file is stored in a centralized location, and it can be modified without being possible to track or even detect the changes. For this reason, in our system, metadata are encoded in a JWT and the base64url representation of the JWT is stored in the contract: this is the return value of the \texttt{tokenURI} method. Therefore, in our system, the metadata extension is used to retrieve the metadata themselves and not a URI that points to the metadata. 

\subsection{Setup}
During setup, resource owners configure their resource servers with $Contract_{AS}$. Moreover, clients ``register'' with the authorization server.\footnote{This registration step is also assumed by the OAuth protocol.} During this registration process, which is out of the scope of this paper, the authorization server learns the Ethereum public key of the client (i.e., $P_{Client}$). In cases where this registration process cannot take place a priori, solutions such as the OAuth~2.0 dynamic client registration protocol~\cite{rfc7591} can be considered.  

Additionally, prior to any other operation, each client obtains from the resource owner an authorization grant (the  format of this grant and the mechanisms for generating it are out of the scope of this paper). This grant represents a resource owner's authorization, and is used by a client to request an access token for a specific $URI_{resource}$. 
 
\subsection{Access token request}
A client, that owns $P_{Client}$, requests from an authorization server an access token for a protected resource $URI_{resource}$, including in the request the authorization grant received during the \emph{setup}. The authorization server verifies $P_{Client}$ ownership and grant validity (using protocols out of the scope of this paper). Then, the authorization server creates a JWT which contains the following claims:
\begin{minipage}{\linewidth}
    \begin{lstlisting} [caption={The generated JWT},label={list:jwt}]
{
    ``iss'': ``$Contract_{AS}$'',
    ``sub'': ``$P_{Client}$'',
    ``aud'': ``$URI_{resource}$''
    ``jti'': ``$token_{id}$''
    ``exp'': ``expiration time''  
}
    \end{lstlisting} 
\end{minipage} 

As a next step, the authorization server creates an ERC-721 token. The $token_{id}$ of this new token matches the value specified by the \texttt{jti} claim of the JWT. Moreover, the metadata of the token is set equal to the base64url encoding of the JWT. Finally, the authorization server invokes the \texttt{transferFrom} method of the ERC-721 contract to transfer the created token to $P_{Client}$, and sends the generated JWT back to the client. It should be noted that the client does not have to store the JWT: at any time it can retrieve all the tokens it owns from the ERC-721 contract, and extract the corresponding JWT from a token's metadata. 

\subsection{Resource access request}
In order for a client to access a protected resource, it sends to the resource server a request that includes the received JWT. The resource server performs the following steps: 
\begin{enumerate}
\item It examines: (i) if it ``knows'' the $Contract_{AS}$  included in the \texttt{iss} claim of the JWT (i.e., if it has been configured with this contract address), (ii) if the $URI_{resource}$ included in the \texttt{aud} claim of the JWT matches the URI of the requested resource, and  (iii) if the token is still valid (i.e., it has not expired).
\item It executes the \texttt{ownerOf} method of $Contract_{AS}$, providing as input the $token_{id}$ included in the \texttt{jti} claim of the JWT, and examines if the returned address corresponds to $P_{Client}$ included in the \texttt{sub} claim of the JWT.
\item It executes the \texttt{tokenURI} method of $Contract_{AS}$, and examines if the returned string is the same as the received JWT.  
\end{enumerate}
At this point, the resource server is able to attest the integrity of the received JWT (since it is the same as the metadata of the token stored in the blockchain), as well as its validity. As a next step, the resource server verifies that the client is the real owner of $P_{Client}$. If all verifications succeed, the resource server accepts the client's request. 

Additionally, a resource server may store a valid JWT locally, create a \emph{session identifier}, and send this identifier to the client: the client may use this identifier, as long as the JWT is still valid, accelerating this way subsequent requests for the same resource. 
 \section{Token management services}
 \label{sec:tok}
 \subsection{Revocation}
 Tokens usually carry an expiration time: OAuth~2.0 and JWTs specifications do not provide any mechanism that allows an authorization server to revoke an access token prior to its expiration time. Even ``OAuth 2.0 token revocation RFC~\cite{rfc7009},'' specifies  a mechanism that allows ``\emph{clients} to notify the authorization server that a [...] access token is no longer needed,'' i.e., a mechanism that provides a ``log out'' functionality rather than revocation.   
 
 In our system, tokens can be revoked by an authorization server by invoking the \texttt{transferFrom} method of the ERC-721 contract and transferring a token back to the authorization server.\footnote{It is reminded that authorization servers are the contract owners, and contract owners in our system are allowed to transfer any token, no matter who the token owner is.} We consider two cases: (i) the corresponding JWT has not been used by the client by the time of the revocation, and (ii) the corresponding JWT has been used, it has been stored locally by the resource server, and it has been associated with a session identifier. In the former case, when a client tries to use the JWT the verification process will fail, since the output of the \texttt{ownerOf} method will not match the $P_{Client}$ included in the \texttt{sub} claim of the JWT, hence the resource server will reject the JTW. In the latter case, the resource server must ``listen'' for the events emitted by the \texttt{transferFrom} method; if an event contains a $token_{id}$ included in an already stored JWT, the resource server must delete the JWT and the associated session identifier. It should be noted that events are immutably stored in the blockchain, hence if a resource server is offline for some time, it can easily recover all missed events. 

 \subsection{Delegation}
 There can be cases where a client does not wish to authenticate to the resource server using $P_{Client}$, e.g., $P_{Client}$ may be stored in a secure, offline storage place, or a user may want to temporary use another device that does not have access to $P_{Client}$ (for example a user may want to use different devices while travelling). For these cases, our system allows a token owner to delegate an access token to another Ethereum address. The token does not change ownership, and the latter address is not allowed to further delegate the token. We implement this functionality by using the \texttt{approve} method of the ERC-721 contract and the ``proof-of-possession key Semantics for JWTs'' defined in RFC 7800~\cite{Jon2016}. RFC 7800 defines a ``confirmation'' (\texttt{cnf}) claim, which contains the key of the owner of a JWT. Delegation of a $token_{id}$ is implemented as follows: $P_{Client}$ invokes the \texttt{approve} method providing as input $token_{id}$ and the Ethereum address associated with the public key of the delegee $P_{Delegee}$. Now $P_{Delegee}$ can use this token by constructing a JWT and by adding $P_{Delegee}$ in the \texttt{cnf} claim as illustrated in listing~\ref{list:jwt}.

 \begin{minipage}{\linewidth}
    \begin{lstlisting} [caption={JWT with delegation enabled. cnf stands for ``confirmation''},label={list:jwt}]
{
    ``iss'': ``$Contract_{AS}$'',
    ``sub'': ``$P_{Client}$'',
    ``aud'': ``$URI_{resource}$''
    ``jti'': ``$token_{id}$''
    ``exp'': ``expiration time''
    ``cnf'':
    {
        ``kid'': ``$P_{Delegee}$''
    }  
}
    \end{lstlisting} 
\end{minipage}

The \texttt{cnf} claim is not recorded in the ERC-721 token's metadata, since metadata are read-only. For this reason, when $P_{Delegee}$ performs a resource access request, Step 3 of token integrity verification is modified as follows. It should be noted that when a token is revoked, the delegee can not use it.

\begin{enumerate}
    \setcounter{enumi}{2}
    \item The resource server executes the \texttt{tokenURI} method of $Contract_{AS}$, and examines if the returned string is the same as the received JWT \textbf{excluding the cnf claim}.  
\end{enumerate}

Then, JWT ownership is verified as follows: the resource server executes the \texttt{getApproved} method of the ERC-721 contract, providing as input $token_{id}$, and examines if the return value equals to the Ethereum address associated with $P_{Delegee}$; finally, it challenges client to verify $P_{Delegee}$ ownership.

The delegation process does not involve the resource owner, neither the authorization server. 
\subsection{Fair exchange}
Smart contracts are ideal for performing ``fair exchange'' of digital goods~\cite{Dzi2018}. In our previous work, published in~\cite{Sir2019}, we used smart contracts to exchange an access token for money. In a nutshell, with the solution presented in~\cite{Sir2019} the authorization server encrypts a token, the client ``deposits'' some money in the form of escrow, and the server receives the escrow only if it reveals the ``correct'' decryption key. The problem with this approach is that it does not provide any guarantee that the decrypted plaintext is a valid access token. 

With the solution presented in this paper this problem is solved as follows: the authorization server creates the ERC-721 token, and stores it in the ERC-721 smart contract by ``indicating'' a $P_{Client}$; the contract transfers the token to $P_{Client}$ only when the latter performs an action (e.g., pay a pre-specified amount of money). The advantage of this approach is that the client can inspect the token before performing any action; of course the client cannot use the token, before performing the specified action, since up until this point, the client does not own the token.      
%

\section{Implementation and Evaluation}
\label{sec:eval}
\subsection{Implementation}
We have developed a proof of concept implementation of the presented solution. We considered the case of an IoT gateway access, the owner of which wishes to grant access to guest users. As an IoT gateway we used Mozilla's WebThings Gateway\footnote{https://iot.mozilla.org/gateway/} that implements the WoT standard~\cite{wot}. For our proof of concept, we chose not to modify the gateway itself, instead, we have developed an application that acts as a proxy, between the client, the blockchain, and the gateway (that holds the role of the resource server).   

As an Ethereum wallet, we used the Metamask Firefox extension, which can handle
ERC-721-based tokens. We implemented clients as JavaScript web applications using web3.js
Ethereum JavaScript API.\footnote{https://web3js.readthedocs.io/}

The main component of our proposed system is the smart contract that implements the functions of the ERC-721 interface. The smart contract was developed using Solidity\footnote{https://solidity.readthedocs.io/en/v0.5.6/},
which is a Turing-complete programming language for implementing smart contracts. In addition to the functions included in Table~\ref{table:erc} we implemented two functions that are not defined in the ERC-721 
interface. The first one, named \texttt{mint}, is for creating new tokens, and the other, named \texttt{burn}, for ``burning'' tokens, i.e., for destroying them. These functions, as well as the \emph{transferFrom} and \emph{approve} functions, can only be invoked by the smart contract owner.

\subsection{Performance and Cost Evaluation}
We have tested our proposed system in the Rinkeby Ethereum test network\footnote{https://www.rinkeby.io/}. We chose a public test network rather than a private one in order to have more reliable results.

The actions of our system that involve the invocation of the smart contract functions, create some computational overhead. Gas is Ethereum's unit for measuring the computational and the storage resources required. Each operation of a smart contract costs a fix amount of gas. Gas cost is the number of units of gas required to perform an action, while gas price is the amount of ``ether'' (i.e., Ethereum's specific coin) a client is willing to pay per unit of gas. The average price of a unit of gas\footnote{As measured by https://ethgasstation.info on 30 Dec. 2019} is \$$0.011 \times 10^{-4}$. Table~\ref{table:ev} shows the cost of deploying the smart contract in the blockchain network, as well as the cost of operations performed by our system in terms of gas units.

\begin{table}[h]
    \centering
    \caption{Cost of our construction building blocks}
    \begin{tabular}{ | l | c |}
    \hline
    \textbf{Operation} & \textbf{Cost measured in gas} \\ \hline
      Contract Deployment  &     1585444                      \\
      Create a token  &         254141                    \\
      Burn a token  &           85791                 \\
      Transfer a token  &      63858                       \\
      Approve           & 45735           \\ \hline
    \end{tabular}
    \label{table:ev}
\end{table}

In addition to gas, Ethereum adds an execution time overhead related to the time the Ethereum network needs to generate a new block. On average, an operation in Ethereum is executed in $\sim$13 seconds.

Some of the aforementioned functions are declared as \emph{view functions}. That is, they only \emph{read} state of the blockchain without modifying it. Thus, they incur no cost, delay, or overhead. These functions are: \texttt{tokenURI}, \texttt{getApproved}, and \texttt{ownerOf}.


\subsection{Discussion}
\textbf{Privacy considerations.} 
With our solution, an authorization server does not have to be aware of the resource server, i.e., they never have to communicate directly, not even in the case of a token revocation. The authorization server learns only the $URI_{resource}$ that a client wants to access, but this does not have to be the real URI of the resource: any pseudonym that the resource server can understand can be used instead. On the other hand, the metadata of an ERC-721 token are immutable and visible to anybody, constituting a privacy threat. In order to enhance clients' privacy metadata can be encrypted using a key known only to the resource owner, to the client, and to the resource server. 

\textbf{Authorization server key breach.} 
In order to enhance the security of our solution we specify that some functions of the ERC-721 contract can only be executed by the contract owner, i.e., the authorization server. Of course, if the private key the authorization server used for deploying the contract is compromised then, the security of our system is jeopardized and a new contract has to be deployed (which requires re-configuration of the resource servers). For this reason, a realization of our system may consider two different keys: one for invoking the security critical functions of the smart contract, and another for specifying which is the former key. The latter key can be the one used for deploying the contract and can be securely stored offline. 

\textbf{Approving other types of identifiers.} ERC-721 specifications define that with the \texttt{approve} method a token owner can specify another Ethereum address that can manage its token. However, in our system a delegee never interacts with the blockchain, therefore the input to \texttt{approve} does not have to be an Ethereum address. Other types of delegee identifiers can be considered, such as legacy public keys, or even contemporary forms of authentication such as verifiable credentials~\cite{ver}.

\section{Conclusions}
\label{sec:conc}
In this paper we presented the design and implementation of an OAuth~2.0 authorization token, backed by blockchain-based smart contracts. Our solution uses Ethereum to store information that can be used for auditing purposes, as well as for token integrity verification. Ethereum smart contracts facilitate revocation, decouple authorization and resource servers, enable token delegation, and create opportunities for exchanging tokens with fungible assets. We believe that our work can be extended towards many directions, including the use of permissioned ledgers, token validation using conditions stored in the ledger, privacy-preserving token representations, and the application of similar mechanisms for managing authorization grants.

\bibliographystyle{IEEEtran}
\bibliography{IEEEabrv,2020-ndss-diss}

\end{document}